\documentclass[preprint,onecolumn,aps,prd,groupedaddress,showpacs,nofootinbib]{revtex4}%

\usepackage{epsfig}
\usepackage{amsmath,amssymb}
\usepackage{amsfonts}
\usepackage{mathrsfs}
\usepackage{relsize}
\RequirePackage{xspace}
\usepackage{graphicx}
\usepackage{slashed}
\usepackage{bm}

\begin{document}

\title{\mbox{}\\[10pt]
Study of Color Octet Matrix Elements Through $J/\psi$ Production in $e^{+}e^{-}$ Annihilation }

\author{ Yi-Jie Li $^{(a)}$}
\author{Guang-Zhi Xu $^{(a)}$}
\author{ Pan-Pan Zhang $^{(a)}$}
\author{Yu-Jie Zhang $^{(b,c)}$}
\email{nophy0@gmail.com}
\author{Kui-Yong Liu $^{(a)}$}
\email{liukuiyong@lnu.edu.cn}
\affiliation{ {\footnotesize (a)~Department of Physics, Liaoning University, Shenyang 110036
, China}\\
  {\footnotesize (b)~School of Physics,
  Beihang University, Beijing 100191, China}\\
  {\footnotesize (c)~CAS Center for Excellence in Particle Physics, Beijing 100049, China}}


\begin{abstract}
In this paper, the color-octet long distance matrix elements are studied though the inclusive $J/\psi$ production in $e^{+}e^{-}$ annihilation within the non-relativistic QCD frame.
The calculations are up-to next-to-leading order with the radiative corrections and relativistic corrections at
the B-factory energy region and the near-threshold region of $4.6\sim5.6~{\rm GeV}$.
A constraint of the long distance matrix elements ($\langle^1S_0^{8}\rangle$,
$\langle^3P_0^{8}\rangle$) is obtained.
Through our estimation, the P-wave color-octet matrix element ( $\langle0|^3P^8_0|0\rangle$ ) should be the order of
$0.005m_c^2~{\rm GeV}^3$  or less.
The constraint region is not compatible with the values of the long distance matrix elements fitted at hadron colliders.
\end{abstract}


\pacs{12.38.Bx,12.39.St,13.85.Ni,14.40.Gx}


\maketitle

\section{Introduction}

The nonrelativistic quantum chromodynamics (NRQCD)\cite{Bodwin:1994jh} factorization approach is widely used to describe the production and decay of heavy quarkonium and
acquires significant achievements since it was proposed.
The surplus production of $J/\psi$ and $\psi^\prime$ in the large transverse momentum region at the Tevatron and the LHC
seems to be a powerful evidence in favor of NRQCD\cite{Affolder:2000nn}.
In particular, the complete next-to-leading order (NLO) corrections (including the contributions from the color-octet and color-singlet channels) were introduced which illustrate an identical set of long distance matrix elements (LDMEs) can satisfy all the $J/\psi$ yield data at hadron colliders\cite{Ma:2010yw,Ma:2010jj,Butenschoen:2010rq,Butenschoen:2011yh}.
Additionally, $\chi_c$ production at hadron colliders are proved to be a successful case for the NLO NRQCD calculations\cite{Ma:2010vd}.
For the long-pending polarization puzzle of $J/\psi$ production, it seems also to be understood naturally using the NLO calculation\cite{Chao:2013cca}.
Subsequently, the NLO analysis is also made for the $\eta_c$ production at hadron colliders and fit the data well using a set of compilable LDMEs with the $J/\psi$ polarization\cite{Han:2014jya,Zhang:2014ybe}.

However, the puzzles are still remained.  The university of the LDMEs is one of the most important things.
The color-octet LDMEs extracted from the $J/\psi$ yield data by the different theoretical groups were incompatible with
each other.
For the $J/\psi$ polarization and $\eta_c$ production, there are also different views in which NLO NRQCD calculations cannot explain the data\cite{Gong:2012ug,Butenschoen:2012px,Butenschoen:2014dra}.
The entirely difference conclusions are caused by the different selections of the LDMEs .
It seems the there color-octet LDMEs ($\langle^3S_1^{8}\rangle$, $\langle^1S_0^{8}\rangle$,
$\langle^3P_0^{8}\rangle$) cannot be determined only by the present hadroproduction data of $J/\psi$ production independently.
In the work of Chao's group, they introduce two linear combination of the there color-octet LDMEs.
Only the two new LDMEs are extracted from the data and the $\chi^2$ will be reduced substantially.
It may imply the higher order corrections should be considered to determine the exact magnitude of the three color-octet LDMEs.
Meanwhile, the analysis combined with the data at other colliders \cite{Butenschoen:2011yh} or the data of other charmonium ($\eta_c$)\cite{Han:2014jya,Butenschoen:2014dra,Zhang:2014ybe} production may help to determine the magnitude of there color-octet LDMEs.

The inclusive $J/\psi$ production in $e^{+}e^{-}$ annihilation is a probe to check the color-octet mechanism and will provide a way to place restrictions on the  color-octet LDMEs. The color-octet channels in the inclusive $J/\psi$ production only involve $^1S_0^8$ and $^3P_J^{8}$ channels.
In the pervious work\cite{Zhang:2009ym}, a constraint of the  color-octet LDMEs was obtained using the data of the inclusive $J/\psi$ production at B-factory.
Meanwhile, some data of the exclusive $J/\psi$ production near the $J/\psi$ mass threshold had been released.
In this paper, we attempt to give a constraint of the color-octet LDMEs within the inclusive $J/\psi$ production in $e^{+}e^{-}$ annihilation including the energy region at B-factory and the near-threshold region.

\section{Numerical results and analysis}

\label{sec:num}

\subsection{B-factory Region}

At B-factory, 
the large gaps between the LO predictions and the experimental data
in the exclusive and inclusive production of $J/\psi$ have been greatly alleviated with the NLO corrections in the color-singlet frame\cite{Gong:2009ng,Gong:2009kp,Ma:2008gq,Zhang:2005cha}.
It leaves little room for the color-octet mechanism.
In the previous work\cite{Zhang:2009ym},  the authors gave a limit superior of the combined color-octet matrix elements ($M_k={\langle0|\mathcal{O}(^1S_0^8)|0\rangle}+k{\langle0|\mathcal{O}(^3P_0^8)|0\rangle}/m_c^2$)
using the NLO results in $\alpha_s$.
In this paper, the relativistic corrections are also considered besides the radiative corrections,
then the constraint will be revised as
\begin{equation}\label{Eq:ME_LIMIT}
M_{3.9\pm0.8}^{(\alpha_s, v^2)}<
\begin{cases}
(2.2\pm0.7)\times10^{-2}~{\rm GeV}^3& \quad \mu=2m_c \\
(2.7\pm0.9)\times10^{-2}~{\rm GeV}^3& \quad \mu=\sqrt{s}/2
\end{cases}
\end{equation}
with the following input parameters
\begin{equation}
m_c=1.5\pm0.1~{\rm GeV},\quad \langle v^2 \rangle=0.2\pm0.1.
\end{equation}
The above matrix element $\langle v^2 \rangle$ is defined as the ratios of the NLO LDMEs in $\mathcal{O}(v^2)$ to the LO LDMEs,
\begin{eqnarray}
 \langle v^2\rangle\equiv\frac{\langle\mathcal{O}^{(2)}\rangle}
 {m_c^2\langle\mathcal{O}^{(0)}\rangle}.
\end{eqnarray}
The strong-coupling constant $\alpha_s$ is evolved by the two-loop formula,
\begin{equation}
\frac{\alpha_s(\mu)}{4\pi}=\frac{1}{\beta_0L}-\frac{\beta_1\ln L}{\beta_0^3L^3},
\end{equation}
where $L=\ln(\mu^2/\Lambda_{QCD}^2)$ with $\Lambda_{QCD}\approx388~{\rm MeV}$, $\beta_0=(11/3)C_A-(4/3)T_f n_f$ and $\beta_1=(34/3)C_A^2-4C_F T_f n_f-(20/3)C_A T_f n_f$ are the one-loop and two-loop coefficients of the QCD beta function, respectively. $\mu$ is the renormalization scale and $n_f$ is the active quark flavors which is set to $3$ for charmonium just as the case in this paper.
In Eq.(\ref{Eq:ME_LIMIT}), we choose two cases for the renormalization scale as $\mu=2m_c$ and $\mu=\sqrt{s}/2$, respectively, to show the uncertainties from renormalization scale.

The values of LDMEs of $J/\psi$  extracted from the experimental $J/\psi/\eta_c$ hadronic production by five theory groups
are shown in Tab.(\ref{tab:matrix})
\cite{Butenschoen:2011yh,Chao:2012iv,Gong:2012ug,Bodwin:2014gia,Zhang:2014ybe}.
These values illustrate the difficulties to determined the color-octet LDMEs.
The series of works by Chao's group imply a wide range of values for $\langle^1S_0^8\rangle$ can satisfy the yield and polarization data of the $J/\psi$ production.
The polarization puzzle can be understood in two different ways:
The contributions from $^3P_0^8$ and $^3 S_1^8$ channels should cancel to each other or be small.
The latter way requests very small values of $\langle^3P_0^8\rangle$ and $\langle^3 S_1^8\rangle$.
The corresponding values of the combined matrix elements are also listed in the table.
Most of the sets of LDMEs do not satisfy the constraint in Eq. (\ref{Eq:ME_LIMIT}) except the set of LDMEs fitted by Butenschoen's group.

{
\begin{table}[!hbp]
 \begin{center}
\caption{\label{tab:matrix}The LDMEs of $J/\psi$  extracted from the experimental $J/\psi/\eta_c$ hadronic production by five theory groups in unit of $10^{-2}~{\rm GeV}^3$
\cite{Butenschoen:2011yh,Chao:2012iv,Gong:2012ug,Bodwin:2014gia,Zhang:2014ybe}.}
\begin{tabular}{c|c|c|c}
\hline
&$\langle0|O^{J/\psi}(^1S_0^8)|0\rangle$&$\frac{\langle0|O^{J/\psi}(^3P_0^8)|0\rangle}{m_c^2}$&$M_{3.9\pm0.8}^{J/\psi}$\\
\hline
Butenschoen, et al. &$4.97\pm0.44$&$-0.716\pm0.089$&$2.2\pm0.8$\\
with feed down&$3.04\pm0.35$&$-0.404\pm0.072$&$1.5\pm0.6$\\
\hline
Chao, et al. ,set1&$8.9\pm0.98$&$0.56\pm0.21$&$11.1\pm0.4$\\
set2&$0$&$2.4$&$9.4\pm1.9$\\
set3&$11$&$0$&$11$\\
\hline
Gong, et al. &$9.7\pm0.9$&$-0.95\pm0.25$&$6.0\pm1.5$\\
\hline
Bodwin, et al. &$9.9\pm2.2$&$0.49\pm0.45$&$11.8\pm2.8$\\
\hline
Zhang, et al. &$0.44\sim1.13$&$1.7\pm0.5$&$7.4\pm2.4$\\
\hline
\end{tabular}
\end{center}
\end{table}}

\subsection{Near-Threshold Region}

Near the $J/\psi$ mass threshold, there are no data for the inclusive $J/\psi$ production.
Meanwhile, the BESIII and Belle collaborations measured the cross sections in the $J/\psi\pi^+\pi^-$ final states at $\sqrt{s}=3.2\sim5.5~{\rm GeV}$\cite{Liu:2013dau,Ablikim:2013mio}.
In their fitting\cite{Liu:2013dau}, the continuum cross sections are sightly above $5~{\rm pb}$ and below $10~{\rm pb}$.
The inclusive cross sections will be larger than that of $J/\psi\pi^+\pi^-$ production by an enhancing factor.
Assuming the factor is two at least, the numerical values of inclusive cross sections will be above $10~{\rm pb}$.

{
\begin{table}[!hbp]
 \begin{center}
\caption{\label{tab:sd_nearthresold}
 The short-distance coefficients of different channels for $J/\psi$ Fock states in the center-of-mass energy region of  $4.6\sim5.6~{\rm GeV}$ are listed in this table.
The uncertainties of the color-octet short-distance coefficients are combined from that of $m_c$ and $\langle v^2 \rangle$.
The strong coupling constant $\alpha_s$ is running with the renormalization scale $\mu$.
In each cell, the number outside the bracket is evolved with $\mu=2m_c$ and $\mu=\sqrt{s}/2$ for inside.
}
\begin{tabular}{c|ccc}
\hline
$\sqrt{s}~({\rm GeV})$&$\hat{\sigma}(^3S_1^1)$&$\hat{\sigma}(^1S_0^8)$&$m_c^2\hat{\sigma}(^3P_J^8)$\\
&${\rm pb}/{\rm GeV}^{3}$&${\rm pb}/{\rm GeV}^{3}$&${\rm pb}/{\rm GeV}^{3}$\\
\hline
4.6&2.0~(2.7)&$386.5^{+90.8}_{-79.6}~(452.3^{+86.1}_{-80.2})$&$7037.1^{+659.1}_{-1034.5}~(8186.8^{+1034.1}_{-1522.7})$\\
4.8&1.9~(2.4)&$344.5^{+73.9}_{-66.9}~(391.7^{+67.4}_{-65.2})$&$5191.5^{+387.9}_{-575.4}~(5880.0^{+635.5}_{-868.8})$\\
5.2&1.7~(2.0)&$269.0^{+53.0}_{-45.3}~(290.8^{+45.8}_{-41.3})$&$3021.3^{+114.4}_{-216.1}~(3261.6^{+232.7}_{-348.4})$\\
5.4&1.6~(1.8)&$238.2^{+45.1}_{-39.2}~(251.9^{+38.0}_{-35.0})$&$2380.6^{+63.4}_{-122.7}~(2516.1^{+148.4}_{-214.1})$\\
5.6&1.5~(1.6)&$211.0^{+52.0}_{-33.5}~(218.8^{+44.8}_{-29.2})$&$1906.3^{+86.3}_{-73.6}~(1975.5^{+34.7}_{-138.9})$\\
\hline
\end{tabular}
\end{center}
\end{table}}

For the theoretical calculations, the reliable predictions cannot be obtained at the region extremely near the threshold.
Therefore, our calculations will concentrate on the energy regions of $4.6\sim5.6~{\rm GeV}$ in which the non-perturbative effects are suppressed.
The short-distance cross sections are shown in Tab.(\ref{tab:sd_nearthresold}).
The relativistic and radiative corrections are included for the color-octet channels.
For the color-singlet channel, we give only the leading order results
but one will find their numerical values are smaller in comparison with the color-octet channels
even the higher corrections are considered
(we argue that the NLO corrections may enhance the color-singlet short-distance cross sections by a factor of about 2).
From the table, one can see that the short-distance cross sections of  $^3P_0^{8}$ channel are three order of magnitude larger
than that of color-singlet channel. The short-distance cross sections of $^1S_0^{8}$ channel are about two order of magnitude lager than that of color-singlet channel.

The total cross sections are sensitive to the selections of the color-octet LDMEs.
The large short-distance coefficients shown in Tab.(\ref{tab:sd_nearthresold}) imply the values of the LDMEs should be small. The negative cross sections may be obtained for the large negative $P$-wave LDMEs. For example, using the LDMEs fitted by Butenschoen's group, the total cross sections in the energy region of $4.6\sim5.6~{\rm GeV}$ will be negative.
For the $^1S_0^8$ LDMEs in order of $\mathcal{O}(0.1~{\rm GeV}^3)$, the contributions from $^1S_0^8$ channel may saturate or overshoot the data.

Next, we try to give a constraint of LDMEs using the above calculations in the energy region of $4.6\sim5.6~{\rm GeV}$.
As mentioned above, the numerical values of the total inclusive cross sections must be larger than that of the cross sections in the process of $J/\psi\pi^+\pi^-$ and may be above $10~{\rm pb}$
as the below expression,
\begin{equation}\label{BESS}
\sigma_{{\rm non}-c\overline{c}}=
\sigma(^1S_0^8)+\sigma(^3P_0^8)+\sigma(^3S_1^1)>10~{\rm pb}.
\end{equation}
Considering the color-singlet matrix element $\langle^3S_1^1\rangle$ is at the order of $\mathcal{O}(1~{\rm GeV^3})$ and referring to Tab.(\ref{tab:sd_nearthresold}), the color-singlet cross sections would be less than $10~{\rm pb}$.
Therefore we consider a looser constraint for the CO LDMEs as,
\begin{equation}\label{lowerlimit}
\sigma(^1S_0^8)+\sigma(^3P_0^8)>0~{\rm pb}.
\end{equation}
But, we note that the color-singlet channels may contribute to the total cross sections for the color-singlet LDME is one or two orders of magnitude larger than the color-octet LDMEs.

\begin{figure}
\includegraphics[width=0.35\textwidth,height=0.35\textwidth]{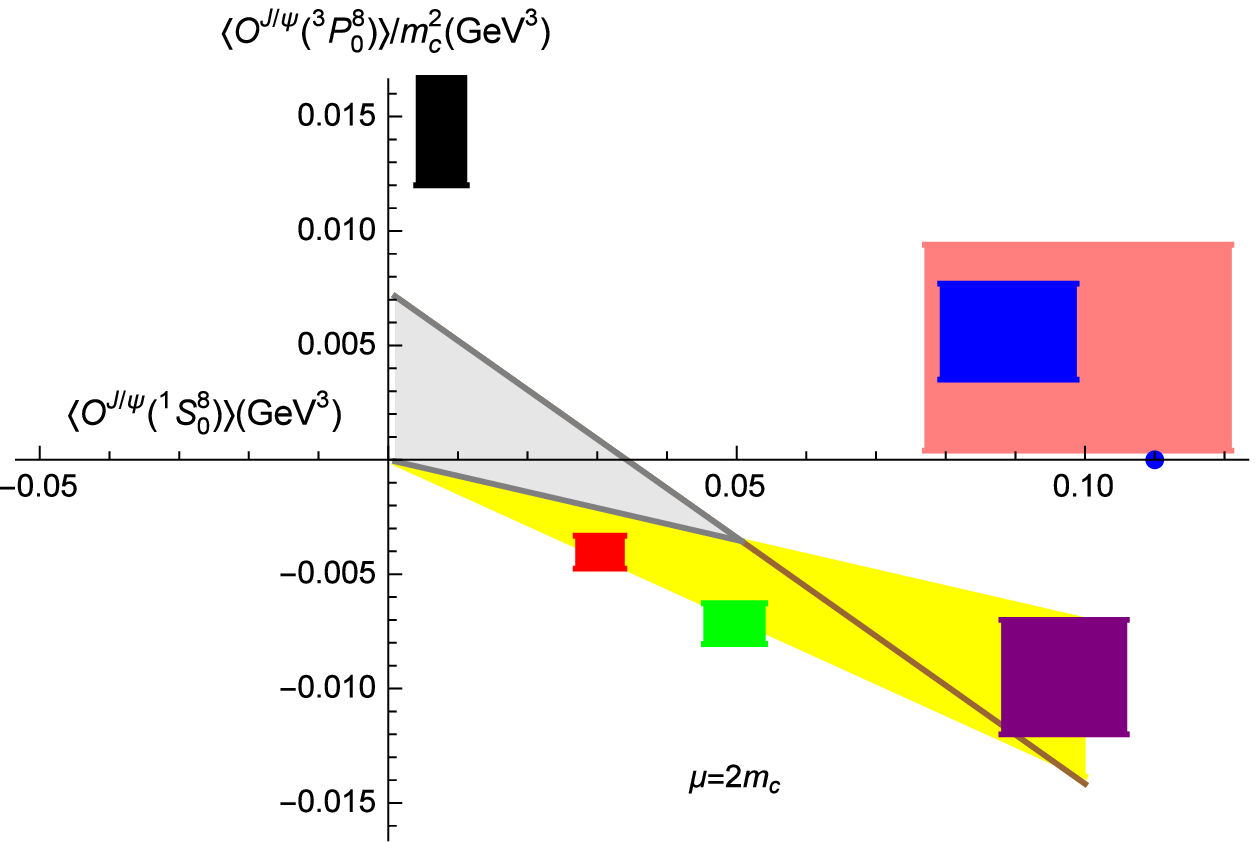}
\includegraphics[width=0.55\textwidth,height=0.35\textwidth]{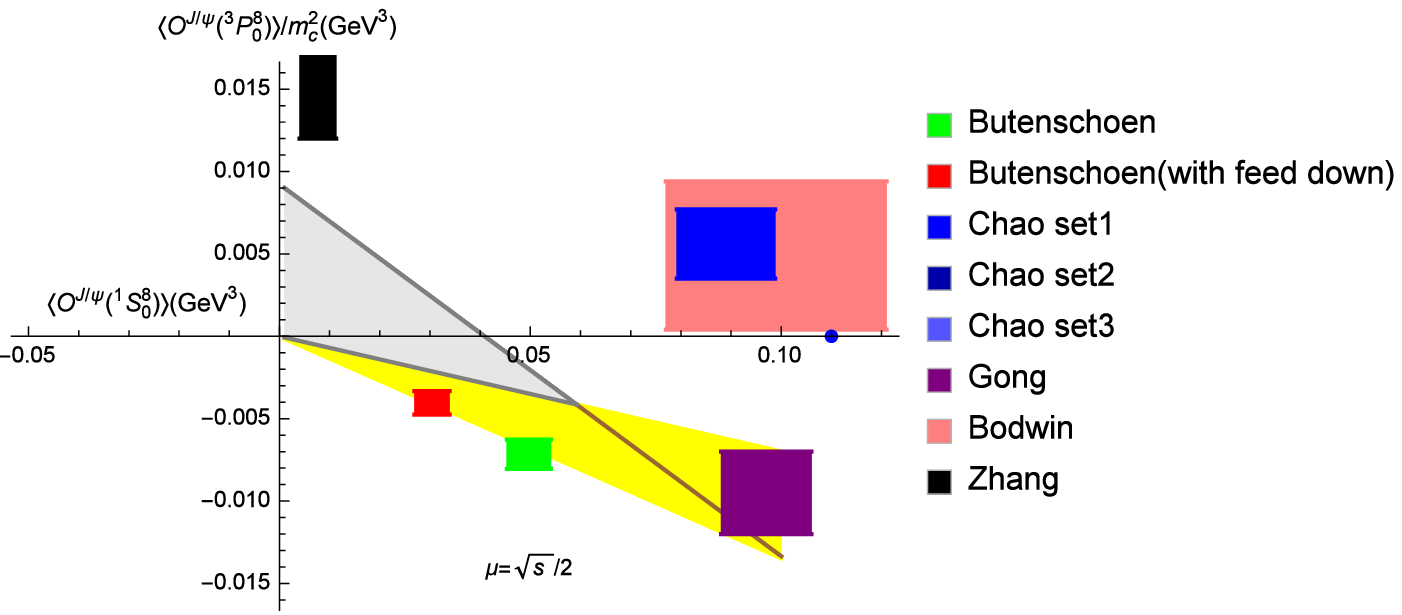}
\caption{ (Color online) The constraints of LDMEs are given using the experimental data of $J/\psi$ inclusive production at B-factory and $J/\psi\pi^+\pi^-$ production in the energy region of $4.6\sim5.6~{\rm GeV}$. The former gives an upper limit corresponding to the brown line. The latter gives a lower limit corresponding to the yellow band.
The gray area presents the combined constraints to LDMEs.
The LDMEs fitting at hadron colliders in Tab.(\ref{tab:matrix}) are also shown in this figure for comparison.}
\label{fig:lems}
\end{figure}

The results of the constraint are shown in Fig.(\ref{fig:lems}).
The gray area shows the constraints to LDMEs combined the data at B-factory and in the energy region of $4.6\sim5.6~{\rm GeV}$.
As shown in Fig.(\ref{fig:lems}) , all the LDMEs fitted at hadron colliders are incompatible with the constraint.
It brings challenges to the university of the NRQCD LDMEs.
The dependence of the renormalization scale can be seen in the figure in which both the cases with $\mu=2m_c$ and $\mu=\sqrt{s}/2$ are given.
The upper limit constrained by B-factory data seems to little heavily depend on the renormalization scale compared with the lower limit by near-threshold data.
The inclusive production in the $e^+e^-$ annihilation restrict the LDMEs to a extremely small area.
The $P$-wave LDMEs $\langle^3P_0^8\rangle$ are restricted into the region of about $-0.005m_c^2\sim0.01m_c^2~{\rm GeV}^3$.
Note that, the inclusive data observed by the Belle and BaBar is not compatible with each other and the upper limit of the constraint region may be different.

\section{Summary}\label{sec:sum}

In this paper, we extend the previous work in Ref.\cite{Zhang:2009ym} and study the color-octet LDMEs ($\langle^1S_0^{8}\rangle$,
$\langle^3P_0^{8}\rangle$) in the inclusive $J/\psi$ production at $e^{+}e^{-}$ colliders.
Both the radiative and relativistic corrections are considered  within NRQCD frame.
An upper limit of the LDMEs is obtained using the data at B-factory.
Refer to the exclusive $J/\psi$ production associated with
$\pi^+\pi^-$ in the energy region of $3.2\sim5.5~{\rm GeV}$.
We argue that the color-octet cross sections of the inclusive $J/\psi$ production must be large than zero picobarn in the region of $4.6\sim5.6~{\rm GeV}$ and a lower limit of the LDMEs is obtained.
Through our estimation, the P-wave color-octet matrix element for $J/\psi$ ( $\langle0|^3P^8_0|0\rangle$ ) should be the order of
$0.005m_c^2~{\rm GeV}^3$  or less.

\begin{acknowledgments}
The authors would like to thank Professor K.T.Chao, C.Meng and Y.Q.Ma for useful
discussion.
This work was
supported by the National Natural Science
Foundation of China (Grants
No. 11375021 and No. 11447018), the New Century Excellent Talents in University (NCET) under grant
NCET-13-0030, the Major State Basic Research Development Program of China (No. 2015CB856701), and General project of Liaoning Provincial Education Department (No. L2015197).
\end{acknowledgments}

\clearpage

\end{document}